\documentclass[twocolumn,aps,prd,10pt,superscriptaddress,longbibliography]{revtex4-1}
\usepackage{amsmath}
\usepackage{amssymb}
\usepackage{bm}
\usepackage{glossaries}
\usepackage{graphicx}
\usepackage{multirow}
\usepackage{fancyvrb}
\usepackage[linesnumbered,ruled]{algorithm2e}
\usepackage{siunitx}
\usepackage{subfigure}
\usepackage{threeparttable}

\usepackage[usenames,dvipsnames,svgnames]{xcolor}
\usepackage{hyperref}
\usepackage[normalem]{ulem}
\usepackage{xcolor}
\usepackage{siunitx}

\hypersetup{
pdfnewwindow=true,      
colorlinks=true,        
linkcolor=Blue,         
citecolor=Blue,         
filecolor=Blue,         
urlcolor=Blue           
}

\newacronym{ace}{ACE}{atomic cluster expansion}
\newacronym{dft}{DFT}{density-functional theory}
\newacronym{dp}{DP}{deep potential}
\newacronym{eam}{EAM}{embedded-atom method}
\newacronym{gap}{GAP}{Gaussian approximation potential}
\newacronym{md}{MD}{molecular dynamics}
\newacronym{mlp}{MLP}{machine-learned potential}
\newacronym{mtp}{MTP}{momentum tensor potential}
\newacronym{nep}{NEP}{neuroevolution potential}
\newacronym{pka}{PKA}{primary knock-on atom}
\newacronym{rmse}{RMSE}{root-mean-square error}
\newacronym{sia}{SIA}{self-interstitial atom}
\newacronym{snes}{SNES}{separable natural evolution strategy}
\newacronym{tabgap}{tabGAP}{tabulated GAP}
\newacronym{zbl}{ZBL}{Ziegler-Biersack-Littmark}

\begin{document}

\title{Large-scale machine-learning molecular dynamics simulation of primary radiation damage in tungsten}

\author{Jiahui Liu}
\affiliation{Beijing Advanced Innovation Center for Materials Genome Engineering, Institute for Advanced Materials and Technology, University of Science and Technology Beijing, Beijing 100083, China}

\author{Jesper Byggm\"{a}star}
\email{jesper.byggmastar@helsinki.fi}
\affiliation{Department of Physics, P.O. Box 43, FI-00014 University of Helsinki, Finland}

\author{Zheyong Fan}
\email{brucenju@gmail.com}
\affiliation{College of Physical Science and Technology, Bohai University, Jinzhou 121013, P. R. China}

\author{Ping Qian}
\affiliation{Beijing Advanced Innovation Center for Materials Genome Engineering, Institute for Advanced Materials and Technology, University of Science and Technology Beijing, Beijing 100083, China}

\author{Yanjing Su}
\email{yjsu@ustb.edu.cn}
\affiliation{Beijing Advanced Innovation Center for Materials Genome Engineering, Institute for Advanced Materials and Technology, University of Science and Technology Beijing, Beijing 100083, China}

\date{\today}

\begin{abstract}

Simulating collision cascades and radiation damage poses a long-standing challenge for existing interatomic potentials, both in terms of accuracy and efficiency. Machine-learning based interatomic potentials have shown sufficiently high accuracy for radiation damage simulations, but most existing ones are still not efficient enough to model high-energy collision cascades with sufficiently large space and time scales. To this end, we here extend the highly efficient neuroevolution potential (NEP) framework by combining it with the Ziegler-Biersack-Littmark (ZBL) screened nuclear repulsion potential, obtaining a NEP-ZBL framework. We train a NEP-ZBL model for tungsten and demonstrate its accuracy in terms of the elastic properties, melting point, and various energetics of defects that are relevant for radiation damage. We then perform large-scale molecular dynamics simulations with the NEP-ZBL model with up to 8.1 million atoms and 240 ps (using a single 40-GB A100 GPU) to study the difference of primary radiation damage in both bulk and thin-foil tungsten. While our findings for bulk tungsten are consistent with existing results simulated by embedded atom method (EAM) models, the radiation damage differs significantly in foils and shows that larger and more vacancy clusters as well as smaller and fewer interstitial clusters are produced due to the presence of a free surface.

\end{abstract}

\maketitle

\section{Introduction}
Fusion reactor materials must be capable of withstanding extremely severe operational conditions \cite{Knaster2016np}. Tungsten (W) is a promising candidate of plasma-facing materials due to its multiple excellent properties such as high melting point, high thermal conductivity and high threshold for sputtering \cite{Gilbert2012nf}. High-energy particles emitted from the fusion plasma initiate collision cascades in the reactor wall material, leading to the formation of lattice defects. Upon later evolution, the created defects cause permanent degradation of the materials such as hardening, swelling, embrittlement, and fracture \cite{Zinkle2013acta}. To ensure a controllable production of fusion energy, it is important to achieve a comprehensive understanding of the structural evolution of the reactor materials under the influence of irradiation \cite{Nordlund2019jnm}.

The generation, distribution and evolution of defects in the early stage of collision cascades are important information for understanding the mechanisms of irradiation resistance \cite{Lin2020acta}. Yi \textit{et al}. observed the presence of dislocation loops with both Burgers vectors $b = 1/2 \left \langle111\right \rangle$ and $b = \left \langle100\right \rangle$ in thin tungsten foils through in-situ self-ion irradiations (150 keV W+) and proposed that the loop nucleation mechanism is likely cascade collapse~\cite{Yi2013pm}. To further investigate the influence of various factors, including irradiation temperature, dose, grain orientation, and material composition, on the radiation damage defect configurations and geometries, the authors conducted a comprehensive in-situ self-ion irradiation study on tungsten and tungsten-based alloys \cite{Yi2015acta, Yi2016acta}. The results showed that dislocation loops with $b = \left \langle100\right \rangle$ and $b = 1/2 \left \langle111\right \rangle$ coexisted in all materials under all irradiation conditions studied, with the majority being of interstitial type. The lowest-dose (0.01 dpa) investigations focused on the “near-surface” cascade effects, revealing that defect clusters formed at individual cascade sites in the form of dislocation loops, most likely of vacancy nature, and of sizes up to $\sim 1300$ point defects. However, this process is beyond the time and length scale of current experiments, making it difficult to experimentally analyze the defect generation and evolution mechanisms at the atomic level. 

Fortunately, the process of primary radiation damage is well within the reach of classical \gls{md} simulations  \cite{Zarkadoul2013jpcm, Nordlund2019jnm, Lin2020acta, Fu2019jnm, Byggmastar2019jpcm, Peng2018nc}. The reliability of \gls{md} simulations, on the other hand, depends crucially on the accuracy of the interatomic potential. For W, a variety of empirical potentials have been developed, but few can accurately describe the various defect structures and material properties related to the collision cascade process, involving self-interstitial clusters \cite{Chen2018jnm}, clustering of vacancies, surface properties \cite{Bonny2014mse}, and local melting followed by rapid recrystallization. In recent years, \glspl{mlp} have shown to be able to accurately describe a variety of physical properties that are relevant for radiation damage in typical materials, including silicon \cite{Hamedani2021prm}, W \cite{Byggmastar2019prb, Granberg2021jnm, Wang2022nf}, aluminium \cite{Wang2019apl}, iron \cite{Wang2022cms, Byggmastar2022jpcm}, and high-entropy alloys \cite{Byggmastar2021prb}. 

A downside of the existing \glspl{mlp} is that they are typically a few orders of magnitude slower than empirical potentials such as the \gls{eam}. Therefore, the \glspl{mlp} developed so far have not been extensively applied to study primary radiation damage in realistically large systems. For example, the \gls{dp} model developed for W \cite{Wang2022nf}, the \gls{mtp} model for Fe \cite{Wang2022cms}, and the \gls{dp} model for Al \cite{Wang2019apl} have not been used to perform large-scale \gls{md} simulations. The \gls{gap} model for W has only been used to anneal structures \cite{Granberg2021jnm} generated by an \gls{eam} potential. The \gls{tabgap} model developed for high-entropy alloys and iron \cite{Byggmastar2021prb,Byggmastar2022prm, Byggmastar2022jpcm} is a notable exception, which is only an order of magnitude slower than \gls{eam} and has been used to perform \gls{md} simulations with up to 3.5 million atoms~\cite{dominguez_nanoindentation_2023}.

Recently, some of the present authors proposed a \gls{mlp} framework called the \gls{nep} \cite{Fan2021prb,Fan2022jcp} that has a computational speed comparable to the \gls{eam} potential, reaching about $2\times 10^7$ atom-step/second in \gls{md} simulations using a single A100 GPU. In this work, we extend the \gls{nep} framework by augmenting it with the \gls{zbl} potential \cite{Ziegler1985} that accounts for the strong repulsion at extremely short interatomic distances. The resulting combined framework, which we call \gls{nep}-\gls{zbl}, retains the high efficiency of the original \gls{nep} framework and at the same time enables the applicability to large-scale radiation damage simulations. 

As an application, we then develop a \gls{nep}-\gls{zbl} model for W using the reference data that has been used for constructing a \gls{gap} model \cite{Byggmastar2019prb}. We evaluate the performance of the \gls{nep}-\gls{zbl} model in terms of elastic constants, melting point, phonon dispersion, and defect energetics that are relevant for radiation damage, with a close comparison with the \gls{gap} model \cite{Byggmastar2019prb} as well as some \gls{eam} potentials. After demonstrating the high accuracy of the \gls{nep}-\gls{zbl} model we then apply it to study primary radiation damage in W using large-scale \gls{md} simulations with up to 8.1 million atoms and 1 million \gls{md} steps. We focus on the differences between irradiation in bulk and thin-foil forms, comparing the generation, distribution and nature of the defects.

\section{Methodologies}

\subsection{The NEP-ZBL framework}

In the \gls{nep} approach \cite{Fan2021prb}, the energy of a system consisting of $N$ atoms can be expressed as the sum of the site energies $U_{i}^{\rm NEP}$ contributed by each atom $i$. The site energy is a function of the atomic-environment descriptor $q^i_{\nu}$. This function is modelled as a feed-forward fully-connected neural network, following Behler and Parrinello \cite{Behler2007prl}. The atomic-environment descriptor $q^i_{\nu}$ consists of a number of radial and angular components, using the Chebyshev and Legendre polynomials up to a certain degree. The name \gls{nep} comes from the training method, which is based on the \gls{snes} \cite{Schaul2011}. For more details on the \gls{nep} approach, the reader is referred to Ref. \onlinecite{Fan2021prb}. 

Accurate characterization of short-range repulsive forces is crucial for simulating early-stage primary radiation damage formation processes. The Ziegler-Biersack-Littmark (ZBL) screened nuclear repulsion potential~\cite{Ziegler1985} has been extensively validated to accurately describe the short-range interactions. In this study, we combine a \gls{nep} model with a repulsive two-body \gls{zbl}-like potential to form a \gls{nep}-\gls{zbl} model. The total site energy $U_i$ on atom $i$ is then
\begin{equation}
U_{i} = U_{i}^{\rm NEP} \left( \{ q_{\nu}^{i} \} \right) 
+ \frac{1}{2} \sum_{j\neq i} U_{\rm ZBL}(r_{ij}).
\end{equation}
Here, we do not use the universal \gls{zbl} potential, but instead the \gls{zbl}-like potential from Ref.~\onlinecite{Byggmastar2019prb} that was specifically optimised for W--W repulsion and used in the GAP model. It has the functional form of the \gls{zbl} potential:
\begin{equation}
U_{\rm ZBL}(r_{ij})=\frac{1}{4\pi\epsilon_{0}}\frac{Z_iZ_je^2}{r_{ij}}\phi(r_{ij}/a)f_{\rm c}(r_{ij}),
\end{equation}
where the screening function is optimised for W--W as~\cite{Byggmastar2019prb}
\begin{align}
\label{phi}
\phi(x) 
&= 0.32825 e^{-2.54931x} \nonumber \\
&+ 0.09219e^{-0.29182x} \nonumber \\
&+ 0.58110e^{-0.59231x},
\end{align}
and 
\begin{equation}
a = \frac{0.46848}{Z_i^{0.23} + Z_j^{0.23}}.
\end{equation}
Here, $\epsilon_0$ is the vacuum dielectric constant, $Z_ie$ is the nuclear charge of atom $i$, and $r_{ij}$ is the distance between atoms $i$ and $j$. For the cutoff function, $f_{\rm c}(r_{ij})$, we take it as the Tersoff one \cite{Tersoff1989prb} with an inner cutoff of 1.0 \AA{} and an outer cutoff of 2.0 \AA. The outer cutoff is significantly shorter than the nearest-neighbour distance in the W lattice, even for self-interstitial configurations. This means that all near-equilibrium properties are left to the \gls{nep} term while the \gls{zbl} term ensures a realistic repulsion when atoms are pushed very close to each other. Note that it is also crucial to ensure that the \gls{nep} term is well-behaved and negligible at very short distances, so that the \gls{zbl} term dominates. This can be achieved by including training structures that contain relatively short interatomic distances, which also ensures a smooth and accurate transition from near-equilibrium distances to the \gls{zbl}-relevant distances. For more discussion on this, see Ref.~\onlinecite{Byggmastar2019prb} and Sec.~\ref{sec:results_training}.

\subsection{Molecular dynamics simulations of high-energy collision cascades}

We apply the \gls{nep}-\gls{zbl} model for W to conduct large-scale \gls{md} simulations to investigate high-energy collision cascades in body-centered cubic (bcc) W. We explore both bulk and thin-foil W systems. All \gls{md} simulations are performed using the \textsc{gpumd} package \cite{Fan2017cpc}.

To prepare the system for initiating a cascade, we equilibrate it under the isothermal-isobaric ensemble for 30 ps, with a target temperature of 300 K and a target pressure of 0 GPa. For foil systems, the free surfaces in the $z$ direction are modeled using open boundaries, while for bulk materials, all three directions are treated as periodic. High-energy particles are created at the center of the simulation box for bulk simulations and near the top surface for thin-foil simulations. Each simulation is run ten times with different selections of the \gls{pka}. The \gls{pka} energies, numbers of simulation steps, box lengths and numbers of atoms, are presented in Table~\ref{table:md_para}. To avoid the channeling effect (requiring prohibitively large simulation cells), the initial momenta of high-energy particles are chosen to be in the high-index direction $\langle135\rangle$. It is essential to acknowledge the potential influence of the incident angle on the formation of defects, e.g., for channeling or near-channeling directions. Consequently, undertaking additional research and conducting comprehensive investigations in the future hold significant value. Atoms within a thickness of $3a_0$ of the boundaries of the simulation boxes (except for the surfaces along the $z$ direction of the thin-foil systems) are maintained at 300 K using the Nose-Hoover chain thermostat \cite{Martyna1992jcp}. The integration time step is dynamically determined so that the fastest atom can move at most 0.01 \AA~ (smaller than 0.5\% of the lattice constant) within one step, with an upper limit of 1 fs also set. Electronic stopping is not considered, as it has not yet been implemented in the \textsc{gpumd} code.

\begin{table}[h]
  \centering \setlength{\tabcolsep}{2mm} 
  \caption{Simulation parameters for bulk and foil systems: the \gls{pka} energy $E_{\rm PKA}$ in units of keV, the number of \gls{md} integration steps $N_t$, the number of bcc unit cells $N_x \times N_y \times N_z$ in the simulation box, and the number of atoms $N$.}
  \begin{tabular}{llllll}
    \hline
    \hline
    & $E_{\rm PKA}$  & $N_t$  & $N_x \times N_y \times N_z$ & $N$\\
    \hline
    bulk & 1   & 50 000  & $30\times30\times30$    & 54 000 \\
         & 5   & 50 000  & $30\times30\times30$    & 54 000 \\
         & 10  & 80 000  & $50\times50\times50$    & 250 000 \\
         & 20  & 100 000 & $80\times80\times80$    & 1 024 000 \\
         & 30  & 100 000 & $100\times100\times100$ & 2 000 000 \\
         & 40  & 100 000 & $100\times100\times100$ & 2 000 000 \\
         & 50  & 100 000 & $100\times100\times100$ & 2 000 000 \\
         & 100 & 200 000 & $120\times120\times120$ & 3 456 000 \\
         & 200 & 200 000 & $150\times150\times150$ & 6 750 000 \\
    \hline
    foil & 100 & 200 000 & $120\times120\times150$ & 4 320 000 \\
         & 200 & 300 000 & $150\times150\times180$ & 8 100 000\\
     \hline
     \hline
    \end{tabular}
      \label{table:md_para}
\end{table}

We used the \textsc{ovito} package \cite{ovito} for defect analyses and visualization. The interstitials and vacancies were identified by using the Wigner-Seitz cell methods \cite{Nordlund1997prb} and the defects were grouped into clusters: two vacancies are considered to be in the same cluster if the distance between them is within the second-nearest-neighbor distance, while the third-nearest-neighbor distance is assumed for self-interstitials. The dislocation analyses were performed using the dislocation extraction algorithm \cite{DXA}. 

\section{Results and discussion}

\subsection{Training a NEP-\gls{zbl} model for W}
\label{sec:results_training}

To train a \gls{nep}-\gls{zbl} model for W, we utilized the training data set for W from Ref.~\onlinecite{Byggmastar2020prm}, which comprises a diverse range of configurations such as dimers, bulk bcc structures with different cell sizes and defects (vacancies, self-interstitials), bcc structures with surfaces, liquid structures, as well as bcc crystals with high-energy short-range interstitials. The latter as well as the repulsive dimers are important to fit the repulsive \gls{nep}-to-\gls{zbl} transition. It should be noted that the data set includes 3,526 structures, but not all the structures in Ref.~\onlinecite{Byggmastar2020prm} were used, as we removed distorted crystals (fcc, hcp, sc, diamond), which was later realized to be unnecessary. Each structure in the data set has a target energy, and some have a target virial tensor, with every atom having three target force components. For detailed information on the quantum-mechanical \gls{dft} calculations for the reference data, please refer to Ref.~\onlinecite{Byggmastar2020prm}.

The \gls{nep}-\gls{zbl} potential was trained using the \textsc{gpumd} package  \cite{Fan2017cpc}. The various hyperparameters in the \gls{nep} model defined in Ref.~\onlinecite{Fan2021prb} are chosen as follows. The cutoff radii for the radial and angular descriptor components are $r_{\rm c}^{\rm R}=6$ \AA~ and $r_{\rm c}^{\rm A}=4$ \AA, respectively. Note that we do not need larger cutoff radius for the radial descriptor components as is required in the case of describing dispersion forces between molecules \cite{dong2023ijhmt}. The Chebyshev polynomial expansion order for the radial and angular descriptor components are $n_{\rm max}^{\rm R}=15$ and $n_{\rm max}^{\rm A}=10$, respectively. The Legendre polynomial expansion order for the angular descriptor components is $l_{\rm max}=4$. The number of neurons in the hidden layer of the neural network is $100$. The weighting factors in the loss function as defined in Ref.~\onlinecite{Fan2021prb} are $\lambda_{1}=\lambda_{2}=0.05$, $\lambda_{\rm e}=\lambda_{\rm f}=1$, and $\lambda_{\rm v}=0.1$. The population size and the number of generations in the \gls{snes} algorithm \cite{Schaul2011} are $N_{\rm pop}=50$ and $N_{\rm gen}=3\times 10^5$. 

\begin{figure}[h] 
\centering 
\includegraphics[width=\columnwidth]{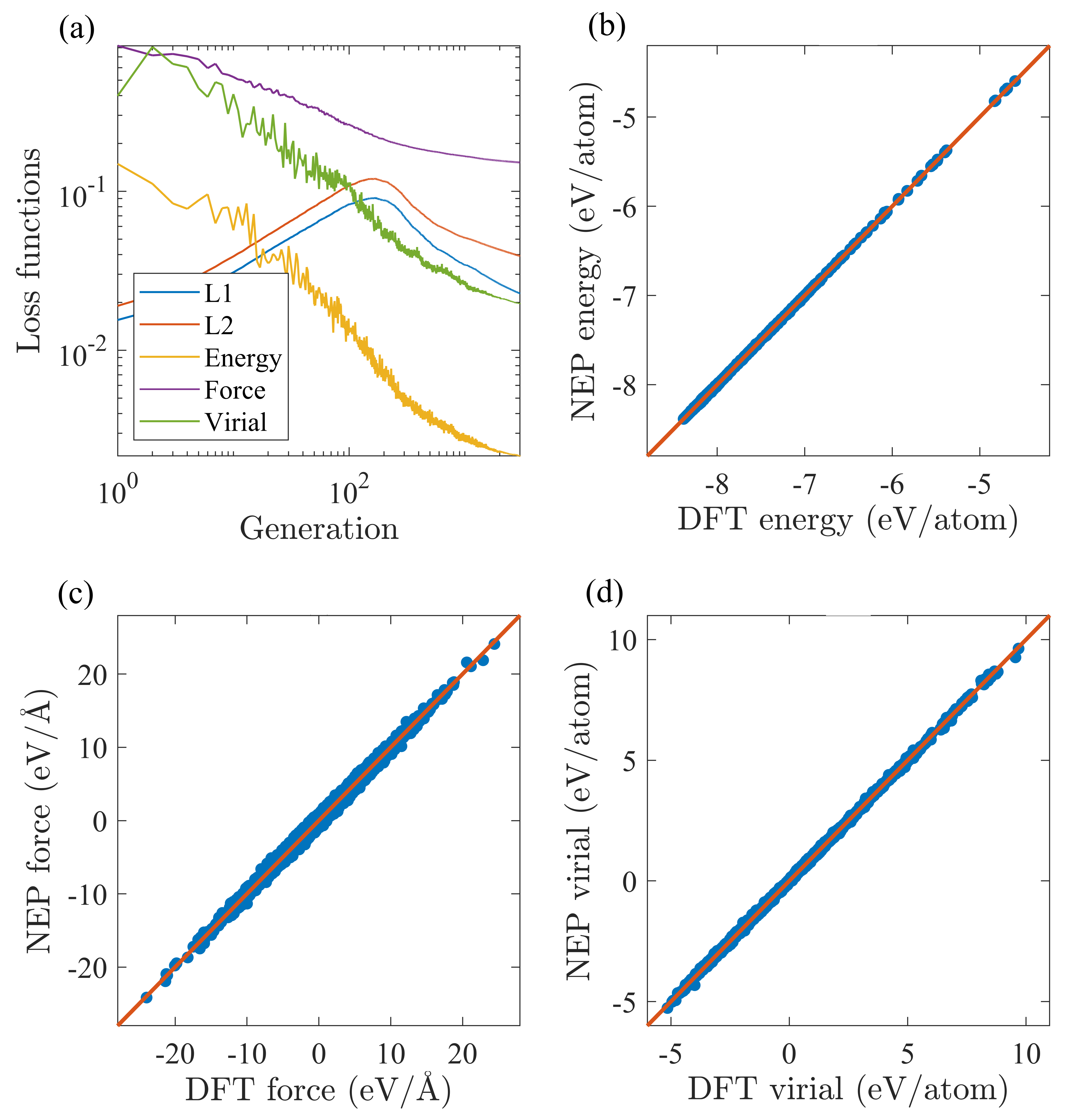}
\caption{(a) Evolution of the various terms in the loss function during the training process, including the $\mathcal{L}_1$ regularization (``L1''), $\mathcal{L}_2$ regularization (``L2''), energy \gls{rmse} (``Energy''), force \gls{rmse} (``Force''), and virial \gls{rmse} (``Virial''). (b) Energy, (c) force, and (d) virial as calculated from the \gls{nep}-\gls{zbl} model compared with the training data, excluding the dimers. } 
\label{fig:nep}
\end{figure}

Figure~\ref{fig:nep}(a) shows the evolution of the various loss terms with respect to the generation. Figure~\ref{fig:nep}(b)-(d) compare the predicted energy, force, and virial values by \gls{nep}-\gls{zbl} and those from quantum-mechanical \gls{dft} calculations for the training set. The \glspl{rmse} of energy, force, and virial for the \gls{nep}-\gls{zbl} model are listed in Table~\ref{table:error}. The accuracy is comparable to that obtained by the \gls{gap}-\gls{zbl} model \cite{Byggmastar2019prb}.

\begin{figure}[h]
\centering
\includegraphics[width=0.75\columnwidth]{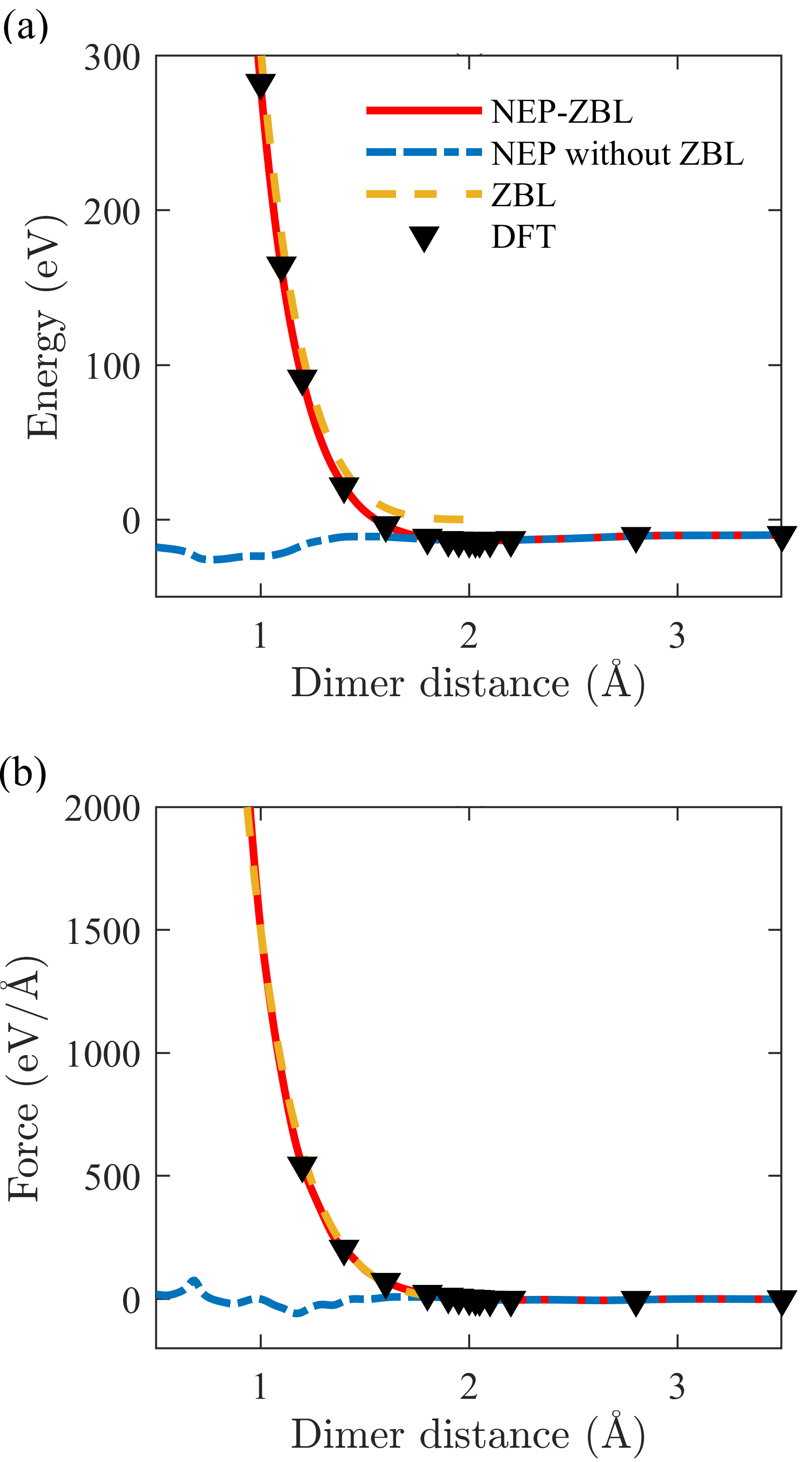}
\caption{(a) Energy and (b) force of short-range repulsion of a W-W dimer calculated by the \gls{nep}-\gls{zbl} model, the \gls{nep} part of \gls{nep}-\gls{zbl}, the \gls{zbl} part of \gls{nep}-\gls{zbl}, and \gls{dft}.}
\label{fig:dimer}
\end{figure}

It is crucial to verify that the \gls{nep}-\gls{zbl} matches the interatomic repulsion provided by the \gls{dft} training data and smoothly connects to and follows the refitted \gls{zbl} potential at short interatomic distances, with no strong or ill-behaved predictions by the \gls{nep} part.  Fig.~\ref{fig:dimer} shows the short-range repulsion of a W-W dimer calculated respectively by the trained \gls{nep}-\gls{zbl}, the \gls{nep} part of \gls{nep}-\gls{zbl}, the \gls{zbl} part of \gls{nep}-\gls{zbl}, and \gls{dft}. The results show that the \gls{nep} part of energy and force at short interatomic distances is of negligible magnitude compared to \gls{zbl} and has little effect on the strong repulsion.

\begin{table}[h] 
\centering \setlength{\tabcolsep}{3mm} 
\caption{Energy \gls{rmse} $\Delta E$, force \gls{rmse} $\Delta F$, and virial \gls{rmse} $\Delta W$ of the training data set from the \gls{nep}-\gls{zbl} model. The corresponding values for the GAP-ZBL model~\cite{Byggmastar2019prb} are given in parentheses. Energy and virial \glspl{rmse} are in units of meV/atom, and force \gls{rmse} is in units of meV/\AA{}. Here, ``Liquid'' include both the liquid structures and the structures with short interatomic distances. ``All'' represents the combination of crystal and liquid structures.} 
\begin{tabular}{lllll} 
\hline \hline 
 & $\Delta E$ & $\Delta F$ & $\Delta W$\\ 
\hline 
Crystal & 1.89 (1.95) & 69.7 (60.0) & 19.6 (14.7) \\ 
Liquid & 5.25 (4.04) & 250 (340) & NA\\ 
All & 2.16 (2.09) & 152 (198) & 19.6 (14.7) \\ 
\hline \hline 
\end{tabular}
\label{table:error}
\end{table}

\subsection{Validating the \gls{nep}-\gls{zbl} model}

\begin{table*}[thb] 
\centering 
\caption{Basic properties of bcc tungsten: cohesive energy per atom~$E_\mathrm{coh}~\rm{(eV/atom)}$, lattice constant~$a~\rm{(\AA)}$, bulk modulus~$B~\rm{(GPa)}$ and elastic constants~$C_{ij}~\rm{(GPa)}$ , (110) surface energy~$E_\mathrm{surf}~\rm{(meV/\AA^2)}$ , formation energies $E_\mathrm{f}$ (eV) of differently oriented self-interstitial dumbbells and mono-vacancy, migration energy of mono-vacancy $E_\mathrm{m}^\mathrm{vac}$ (eV), and melting temperature~$T_\mathrm{melt}~\rm{(K)}$.}
\begin{threeparttable}
\begin{tabular}{llllllllll} 
\hline 
\hline 
&Exp.&DFT&\gls{nep}-\gls{zbl}&GAP \cite{Byggmastar2019prb}&CHEN \cite{Chen2018jnm}&AT \cite{Chen2018jnm}&JW \cite{Chen2018jnm}&MV2-B \cite{Marinica2013jpcm}&MV4-S \cite{Marinica2013jpcm}\\ 
\hline 
$E_\mathrm{coh}$ &-8.9\tnote{a}&-8.39\tnote{d}&-8.38&-8.39&-8.9&-8.9&-8.9&-8.9&-8.9\\
$a$ &3.165\tnote{a}&3.185\tnote{d}&3.185&3.185&3.165&3.165&3.165&3.140&3.143\\
$B$&310\tnote{a}&304\tnote{d}&307&309&310&310&310&320&309\\
$C_{11}$&522\tnote{a}&522\tnote{d}&518&526&522&522&522&544&523\\
$C_{12}$&204\tnote{a}&195\tnote{d}&201&200&204&204&204&208&202\\
$C_{44}$&161\tnote{a}&148\tnote{d}&144&149&161&161&161&160&161\\
$E_\mathrm{surf}$&187\tnote{b}, 203\tnote{b}&204\tnote{d}&205&204&159&161&161&144\tnote{j}&157\tnote{k}\\
$E^{\left \langle 111 \right \rangle \mathrm{db}}_\mathrm{f}$&...&9.55\tnote{e}, 10.29\tnote{f}&10.87&10.38&9.46&8.92&9.50&10.52&10.53\\
$E^{\left \langle 110 \right \rangle \mathrm{db}}_\mathrm{f}$&...&9.84\tnote{e}, 10.58\tnote{f}&11.09&10.59&9.80&9.64&10.16&10.82&10.82\\
$E^{\left \langle 100 \right \rangle \mathrm{db}}_\mathrm{f}$&...&11.49\tnote{e}, 12.20\tnote{f}&12.15&12.11&11.01&9.82&10.30&12.86&12.72\\
$E_\mathrm{f}^\mathrm{vac}$ &$3.67\pm0.2$\tnote{c}& 3.22\tnote{d}&3.28&3.32&3.54&3.63&3.63&3.49&3.81\\
$E_\mathrm{m}^\mathrm{vac}$ &$1.78\pm0.1$\tnote{c}& 1.73\tnote{g}&1.74&1.71&1.91&1.44&1.44&1.85\tnote{j}&1.84\tnote{l}\\
$T_\mathrm{melt}$ & 3687\tnote{a}&3450$\pm$100\tnote{h} & 3540$\pm$10 &3540$\pm$10&4580$\pm$10&5200$\pm$50\tnote{i} &...&...&...\\
\hline 
\hline 
\end{tabular}
\begin{tablenotes}
  \item \tnote{a} Ref.~\onlinecite{Rumble2019crc}, \tnote{b} Ref.~\onlinecite{Tyson1977ss}, \tnote{c} Ref.~\onlinecite{Rasch1980pm}, \tnote{d} Ref.~\onlinecite{Byggmastar2019prb}, \tnote{e} Ref.~\onlinecite{Dudarev2006prb}, \tnote{f} Ref.~\onlinecite{Ma2019prm1},\\ 
  \item \tnote{g} Ref.~\onlinecite{Ma2019prm2}, \tnote{h} Ref.~\onlinecite{Wang2011prb}, \tnote{i} Ref.~\onlinecite{Fikar2009jnm}, \tnote{j} Ref.~\onlinecite{Chen2018jnm}, \tnote{k} Ref.~\onlinecite{Bonny2014mse}, \tnote{l} Ref.~\onlinecite{Liu2020tungsten}.
\end{tablenotes}
\end{threeparttable}
\label{table:property}
\end{table*}

\begin{figure}[h]
\centering
\includegraphics[width=0.85\columnwidth]{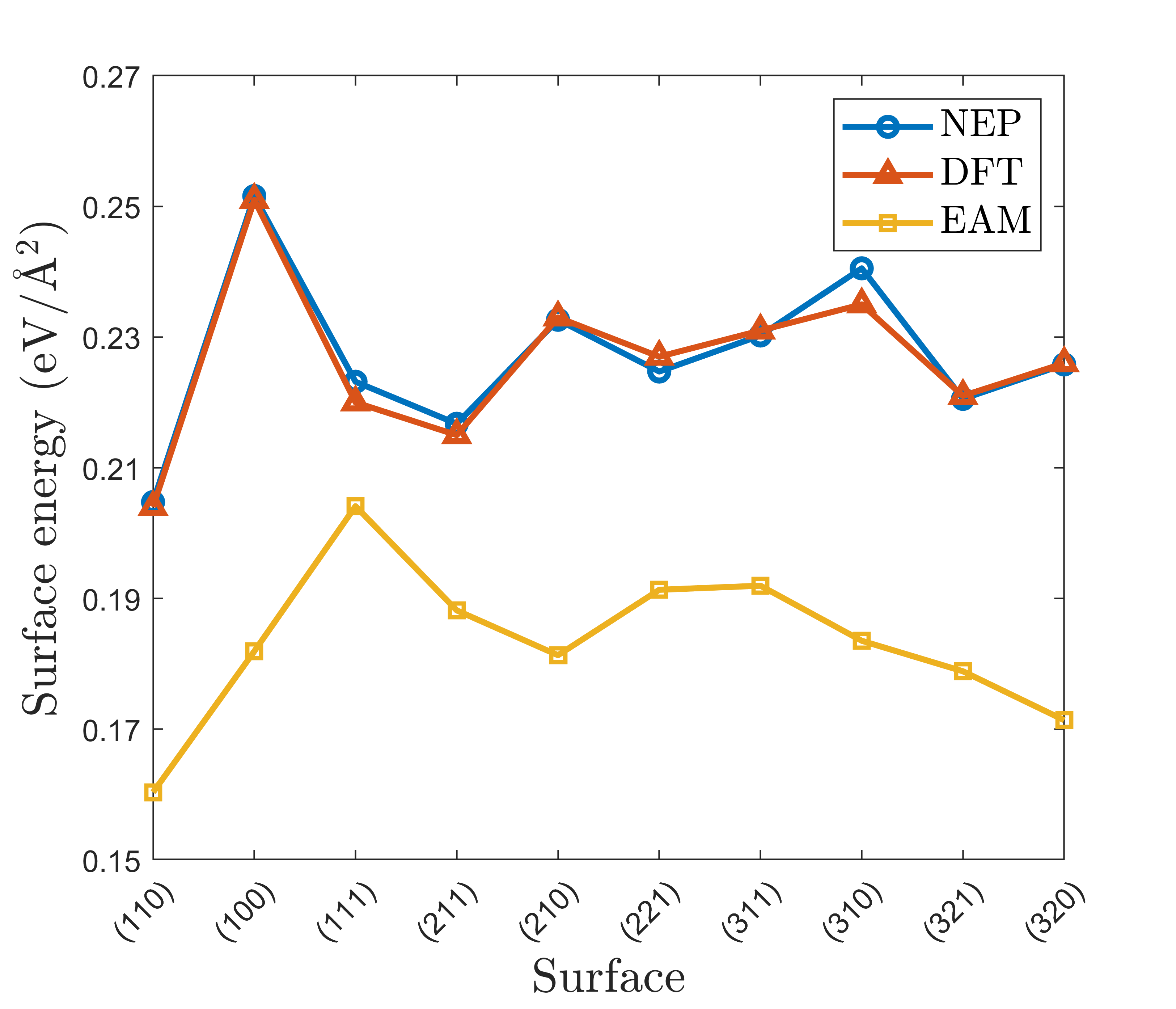}
\caption{Surface energies predicted by the \gls{nep}-\gls{zbl} model as compared with \gls{dft} and \gls{eam} \cite{AT1987pma}. The \gls{dft} values are from Ref. \cite{Byggmastar2019prb}.}
\label{fig:surf}
\end{figure}

To evaluate the reliability of the \gls{nep}-\gls{zbl} model in modelling irradiation effects, we calculated a set of relevant material properties. Static calculations were carried out using \textsc{ase} \cite{Larsen2017jpcm}. In Table~\ref{table:property}, the \gls{nep}-\gls{zbl} results are compared with those from experiments, \gls{dft} calculations, and some empirical potentials. The \gls{nep}-\gls{zbl} model shows a satisfactory agreement in the predictions of lattice parameters, cohesive energies, and elastic constants. As shown in Figure~\ref{fig:surf}, the formation energies of 10 free surfaces calculated by the \gls{nep}-\gls{zbl} model are significantly better than traditional analytical potentials, which often significantly underestimate surface energies and predict the wrong order of stability of different surface orientations~\cite{Bonny2014mse}. Note that only the first four low-index surfaces in Fig.~\ref{fig:surf} are included in the training database, thus the accuracy is evidence of good generalizability outside of the training data. The generalization of surface formation energy can also be achieved by \gls{gap}  \cite{Byggmastar2019prb} and \gls{dp} models \cite{Wang2022nf}.

The point defect formation energies were evaluated in $4 \times 5 \times 6$ supercells for comparing with \gls{dft}. The formation energies of \glspl{sia} calculated by the \gls{nep}-\gls{zbl} model are overestimated compared to the \gls{dft} reference due to the slightly underestimated cohesive energy, but it correctly predicts the relative stability between the interstitial structures. The vacancy formation energy and the vacancy migration barrier are also consistent with \gls{dft}. The binding of divacancies is a peculiar feature of tungsten and some other bcc transition metals \cite{Byggmastar2019prb}. It is reported that the binding energy of the second-nearest neighbor (2NN) divacancy is strongly repulsive ($E_\mathrm{b}(\rm{2NN})=-0.286$ eV \cite{Mason2017jpcm}). In contrast, the interaction between the first-nearest-neighbor (1NN) vacancies is weakly binding ($E_\mathrm{b}(\rm{1NN})=0.048$ eV \cite{Mason2017jpcm} or weakly repulsive ($E_\mathrm{b}(\rm{1NN})=-0.1$ eV~\cite{Becquart2007nimprsb}) depending on \gls{dft} code and settings. In the \gls{nep}-\gls{zbl} model, $E_\mathrm{b}(\rm{1NN})$ is $0.17$ eV and $E_\mathrm{b}(\rm{2NN})$ is $-0.20$ eV. Overall, the \gls{nep}-\gls{zbl} model predicts qualitatively the correct binding energies in good agreement with \gls{dft}. 

\begin{figure}[h]
\centering
\includegraphics[width=0.8\columnwidth]{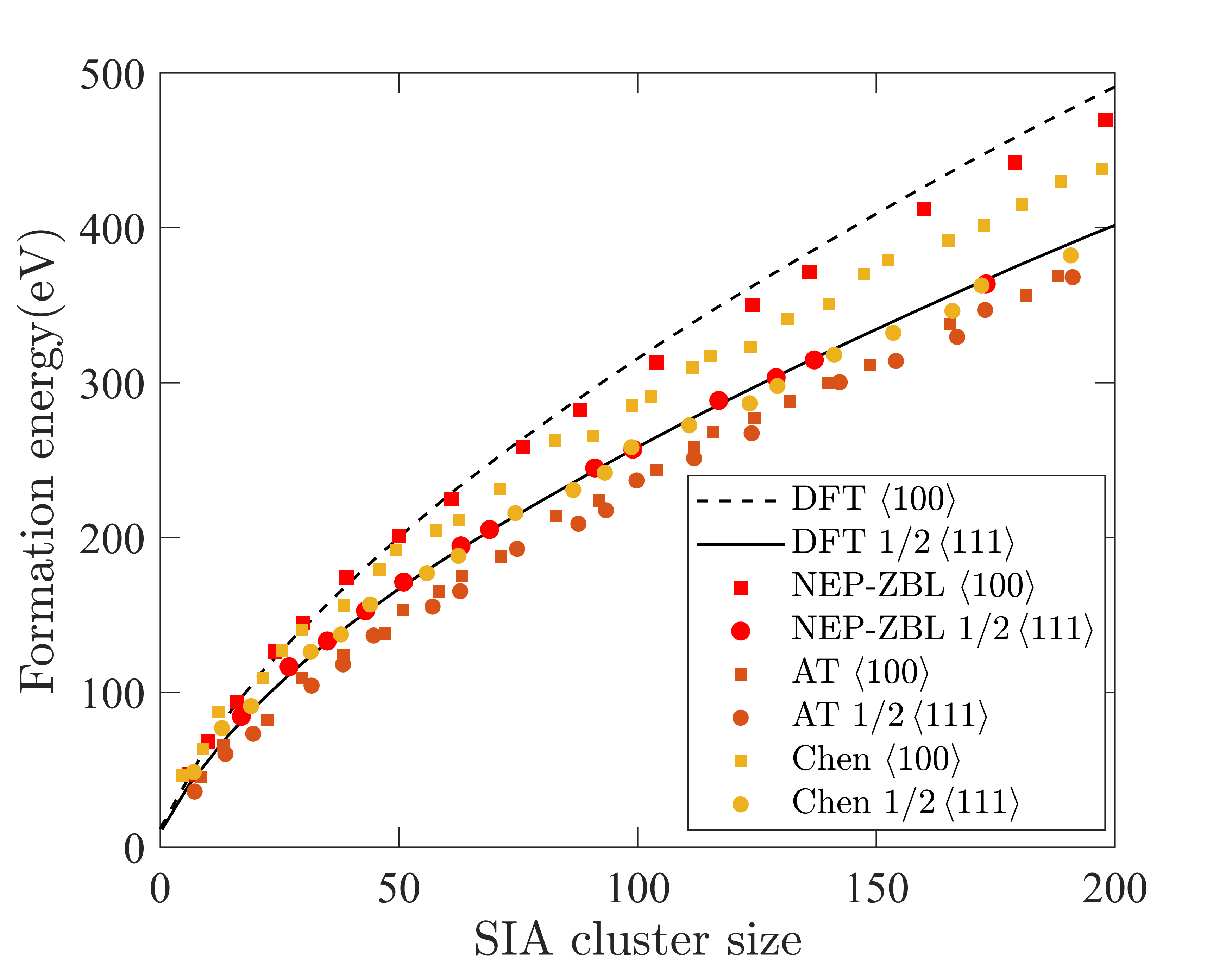}
\caption{Formation energies of $1/2\left \langle111\right \rangle$, $\left \langle100\right \rangle$ clusters in W predicted by \gls{nep}-\gls{zbl}, \gls{dft}, {\color{red}\sout{\gls{eam},}} AT \cite{AT1987pma}, and Chen \cite{Chen2018jnm} data.}
\label{fig:cluster}
\end{figure}

The impact of point defects on the material's mechanical properties is relatively weak compared to the clusters formed due to migration and grouping of vacancies and self-interstitials \cite{Wang2022nf}. Hence, we investigate the formation energy and relative stability of \gls{sia} $1/2\left \langle 111\right \rangle$ and $\left \langle 100\right \rangle$ clusters (dislocation loops) with a box of 250,000 atoms. Figure~\ref{fig:cluster} shows the formation energies of $1/2\left \langle111\right \rangle$ and $\left \langle100\right \rangle$ prismatic loops predicted by \gls{nep}-\gls{zbl}, other \gls{eam} models \cite{Chen2018jnm,AT1987pma} and the \gls{dft} extrapolation model \cite{Alexander2016prb}. The \gls{nep}-\gls{zbl} data closely follows the \gls{dft} extrapolation model, unlike the EAM potentials. 

We also simulated threshold displacement energies ($E_\mathrm{d}$) with the \gls{nep}-\gls{zbl} model at 30 K. The simulation box was a $12 \times 12 \times 16$ supercell containing 4608 atoms. The simulations were cooled down by one lattice atomic layer at boundaries. The time-step was variable and limited the displacement of the fastest atom to 0.005~\AA. The minimum $E_\mathrm{d}$ value is 49 eV for the $\left \langle 100 \right \rangle$ direction. We calculated $E_\mathrm{d}$ for six other directions ($\langle 110 \rangle$, $\langle 122 \rangle$, $\langle 133 \rangle$, $\langle 135 \rangle$, $\langle 235 \rangle$, and $\langle 111 \rangle$) and determined the average $E_\mathrm{d}$ of our potential to be 117 eV. Note that a reliable global average over crystal directions typically requires on the order of a hundred or thousand directions, but this value serves as a first approximation.

\begin{figure}[h]
\centering
\includegraphics[width=0.8\columnwidth]{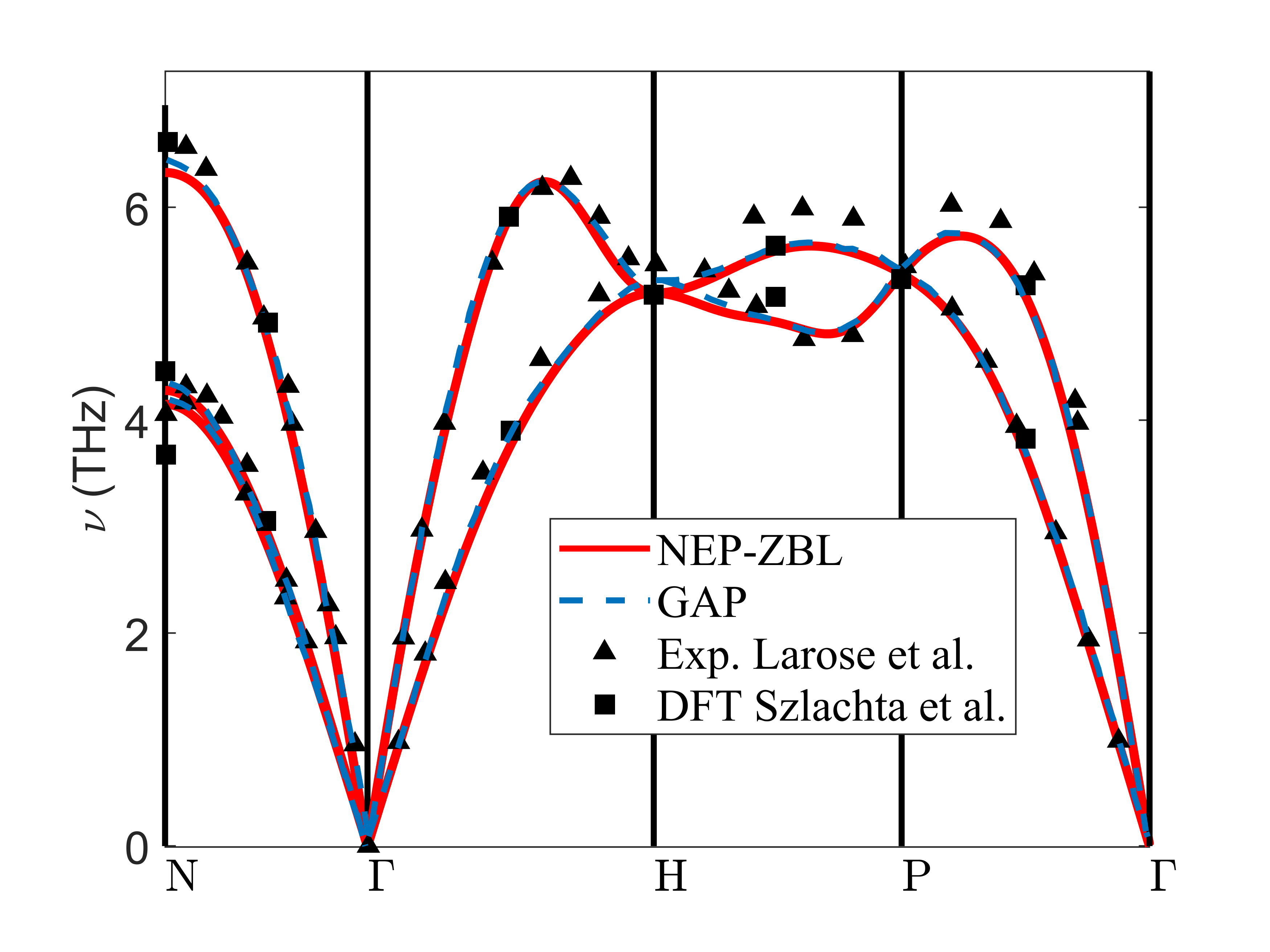}
\caption{Phonon dispersion of bcc W as given by the \gls{nep}-\gls{zbl} and compared with \gls{dft}~\cite{Szlachta2014prb} and experimental data~\cite{Larose1976cjp}.}
\label{fig:phonon}
\end{figure}

The melting point was calculated by the solid-liquid coexistence method~\cite{luo2004jcp}. The bi-phase system containing 13500 atoms with half of the atoms in the liquid phase and the other half in the solid bcc phase was simulated at temperatures ranging from 3500 K to 3600 K, and the pressures were kept at 0 GPa. The system remained as bi-phase using \gls{nep}-\gls{zbl} potential at 3540 K, above which the liquid phase grows, and below this temperature the system crystallizes. Figure~\ref{fig:phonon} presents the phonon dispersion of bcc tungsten calculated by \gls{nep}-\gls{zbl} compared with experimental data~\cite{Larose1976cjp}, the results calculated by GAP \cite{Byggmastar2019prb}, and previous DFT studies~\cite{Szlachta2014prb}. The results show that the dispersion relation is well-reproduced by the \gls{nep}-\gls{zbl} model. All the above results demonstrate that the \gls{nep}-\gls{zbl} potential is not negatively affected by the \gls{zbl} potential and can provide accurate predictions about the material properties near the equilibrium state.

\begin{figure}[h]
\centering
\includegraphics[width=0.8\columnwidth]{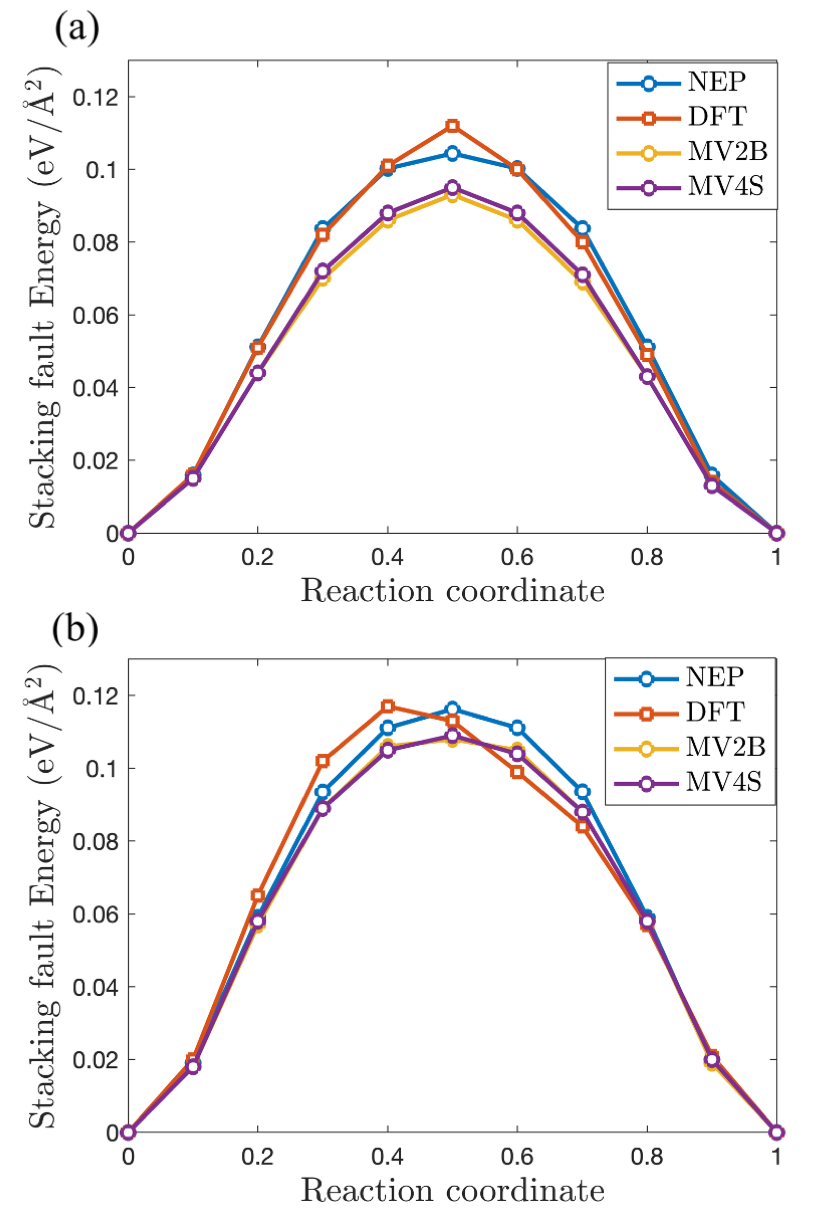}
\caption{Prediction of GSF energy by \gls{dft}~\cite{Wang2022nf}, \gls{nep}-\gls{zbl}, MV2-B~\cite{Marinica2013jpcm} and MV4-S~\cite{Marinica2013jpcm}, along the $\left \langle 111 \right \rangle$ direction on (a) $(1 \Bar{1} 0)$ and (b) $(1 \Bar{1} 2)$ plane.}
\label{fig:stacking_fault}
\end{figure}

The energy landscape of generalized stacking fault (GSF) is defined by the energy change that occur when one section of the crystal is displaced in relation to the other along a specific plane ($\gamma$ plane). The GSF energy along a given direction is referred to as the $\gamma$-line. We evaluate the two common $\gamma$-lines that are relevant for screw dislocation motion using the NEP-ZBL model: displacement along the $\left \langle 111 \right \rangle$ direction for both the $(1 \Bar{1} 0)$ $\gamma$ plane and $(1 \Bar{1} 2)$ $\gamma$ plane. We compare our findings with the results~\cite{Wang2022nf} of $\gamma$-lines computed via \gls{dft} and \gls{eam} models, as depicted in Figure~\ref{fig:stacking_fault}. The predictions of the $\gamma$-lines by \gls{nep}-\gls{zbl} model exhibit a good agreement with \gls{dft}. The \gls{eam} predictions for $\gamma$-lines are marginally lower than \gls{dft} values at both the $(1 \Bar{1} 0)$ and $(1 \Bar{1} 2)$ $\gamma$ planes.

\begin{figure}[h]
\centering
\includegraphics[width=0.8\columnwidth]{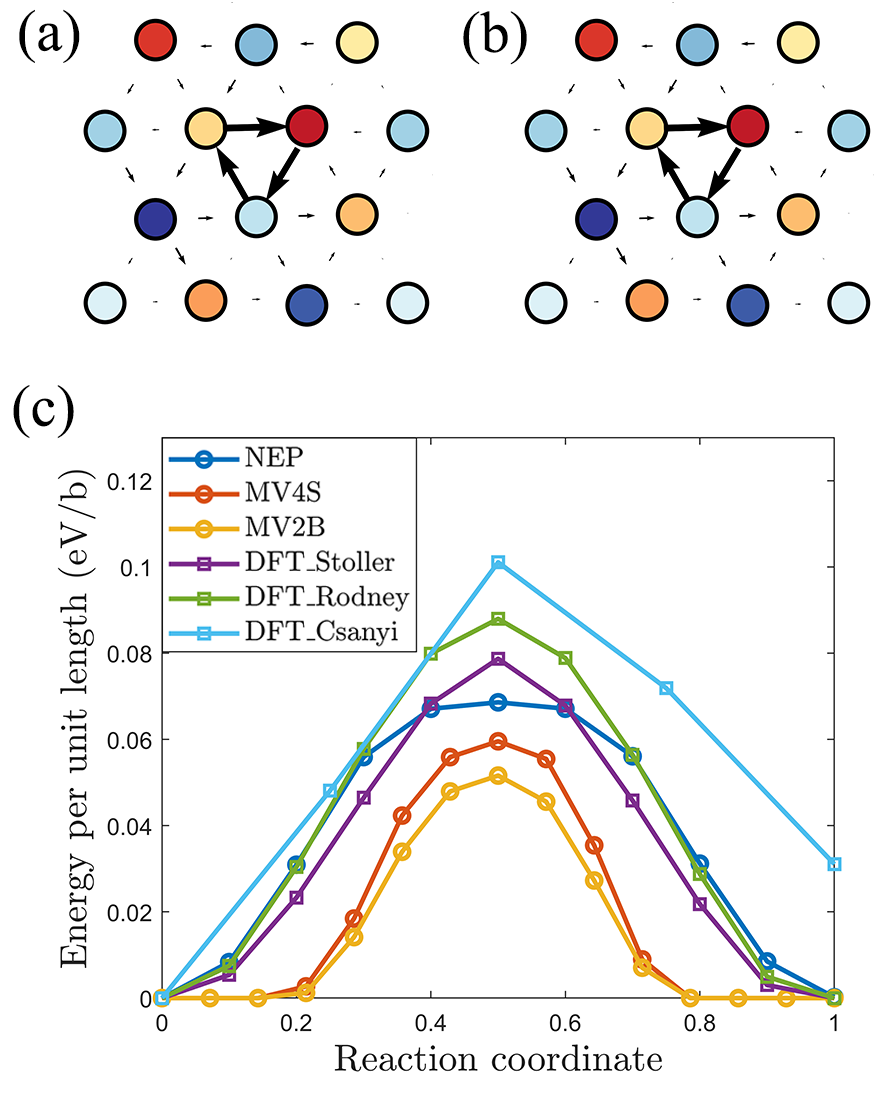}
\caption{Relaxed core structure of a 1/2$\left \langle 111 \right \rangle$ screw dislocation in the (a) NEP-ZBL and (b) DFT simulations, visualised using differential displacement plots~\cite{Vitek1970pm}. The colours represent the different (111) layers spanning one Burgers vector length. Arrows between nearest neighbors indicate the out-of-plane $\left \langle 111 \right \rangle$ displacements with respect to the perfect bulk. (c) Peierls barrier of bcc W as given by the \gls{nep}-\gls{zbl}, MV2-B~\cite{Marinica2013jpcm} and MV4-S~\cite{Marinica2013jpcm}, and compared with \gls{dft} calculated Peierls barriers by Rodney et al~\cite{Ventelon2013acta}, Csanyi et al~\cite{Szlachta2014prb} and Stoller~\cite{Samolyuk2013jpcm}.}
\label{fig:screw}
\end{figure}

Reproducing the fundamental properties of screw dislocations has often been challenging for traditional interatomic potentials. To evaluate the relaxed core of the screw dislocation and Peierls barrier for screw dislocation migration, we employed 135-atom boxes with quadrupolar periodic arrangements of screw dislocation dipoles to generate two screw dislocations with opposite Burgers vectors, following the method established in Ref.~\cite{Ventelon2013acta}. Figure~\ref{fig:screw} presents the relaxed core of the screw dislocation in \gls{nep}-\gls{zbl}~(a) and DFT~(b), while Figure~\ref{fig:screw}~(c) shows the Peierls barrier obtained from the \gls{nep}-\gls{zbl} model, \gls{eam}~\cite{Marinica2013jpcm} models (MV2-B and MV4-S model) and \gls{dft}~\cite{Wang2022nf} calculations. Barriers are determined with simultaneous migration of both dislocations using the NEB method~\cite{NEB_method}. The \gls{nep}-\gls{zbl} model successfully replicates the symmetric nondegenerate core structure of the $1/2\left \langle 111 \right \rangle$ screw dislocation, as predicted by \gls{dft}. Moreover, the \gls{nep}-\gls{zbl} model yields similar barriers with consistent shapes in comparison to the DFT results, demonstrating a notable improvement over the EAM models.

Apart from accuracy, computational efficiency is also vital for simulating primary radiation damage. Table.~\ref{table:efficient} compares \gls{nep}-\gls{zbl} against an \gls{eam}-\gls{zbl} model ~\cite{Chen2018jnm} and a \gls{dp}-\gls{zbl} model ~\cite{Wang2022nf} in terms of the computational speed and the upper limit of the system size in \gls{md} simulations, using one 40-GB A100 GPU. The \gls{nep}-\gls{zbl} model is only a couple of times slower than the empirical \gls{eam}-\gls{zbl} model and is about two orders of magnitude as efficient as the \gls{dp}-\gls{zbl} model. The superior computational efficiency for \gls{nep} aginst \gls{mtp} and \gls{gap} has been discussed previously ~\cite{Fan2022jcp}.

\begin{table}[h] 
\centering \setlength{\tabcolsep}{6mm} 
\caption{The computational speed (in units of atom-step/second) and the maximum system size ($N_{\rm max}$) for \gls{nep}-\gls{zbl}, \gls{eam}-\gls{zbl}~\cite{Chen2018jnm}, and \gls{dp}-\gls{zbl}~\cite{Wang2022nf} as measured using one 40-GB A100 GPU.}
\begin{tabular}{llll}
\hline \hline 
\multicolumn{2}{l}{Potential} & Speed & $N_{\rm max}$ \\
\hline 
\multicolumn{2}{l}{ \gls{eam}-\gls{zbl}}& $3.6\times10^7$ & $9.8\times10^6$\\ 
\multicolumn{2}{l}{ \gls{nep}-\gls{zbl}}& $1.3\times10^7$ & $8.1\times10^6$\\ 
\multicolumn{2}{l}{\gls{dp}-\gls{zbl}}& $1.3\times10^5$ & $4.4\times10^4$\\ 
\hline \hline 
\end{tabular}
\label{table:efficient}
\end{table}

\begin{figure}[h]
\centering
\includegraphics[width=0.8\columnwidth]{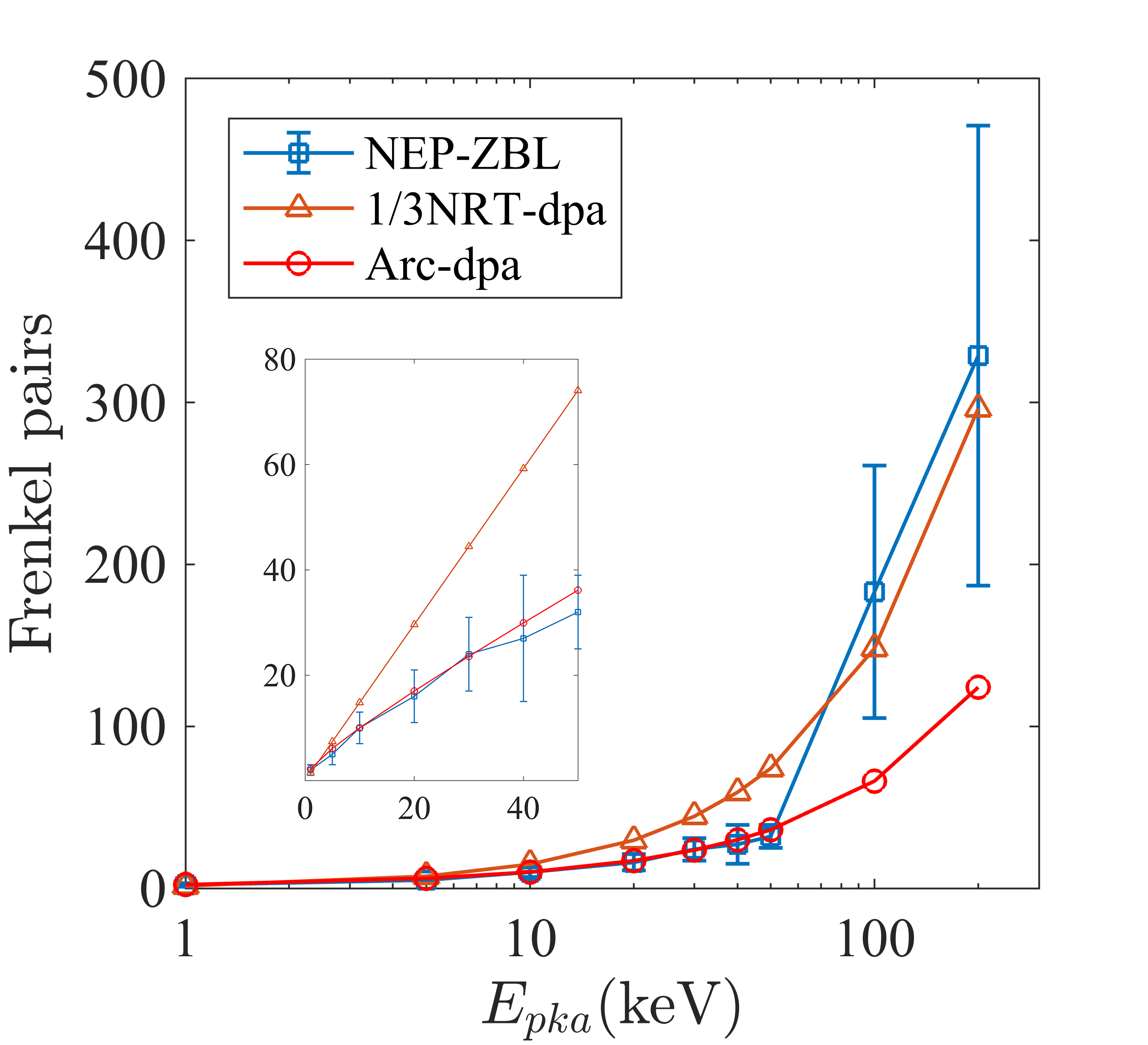}
\caption{The residual point defects, the corresponding 1/3
NRT model results, and the corresponding arc-DPA model results. Each point is the average of 10 independent cascade simulations, and the errors are given in the standard deviation. The inset in the figure shows a duplicate view of the low-value data for better visibility and comparison.}
\label{fig:fks}
\end{figure}

\subsection{Radiation-induced defects in bulk tungsten}

\begin{figure*}[htb]
\centering
\includegraphics[scale= 0.13]{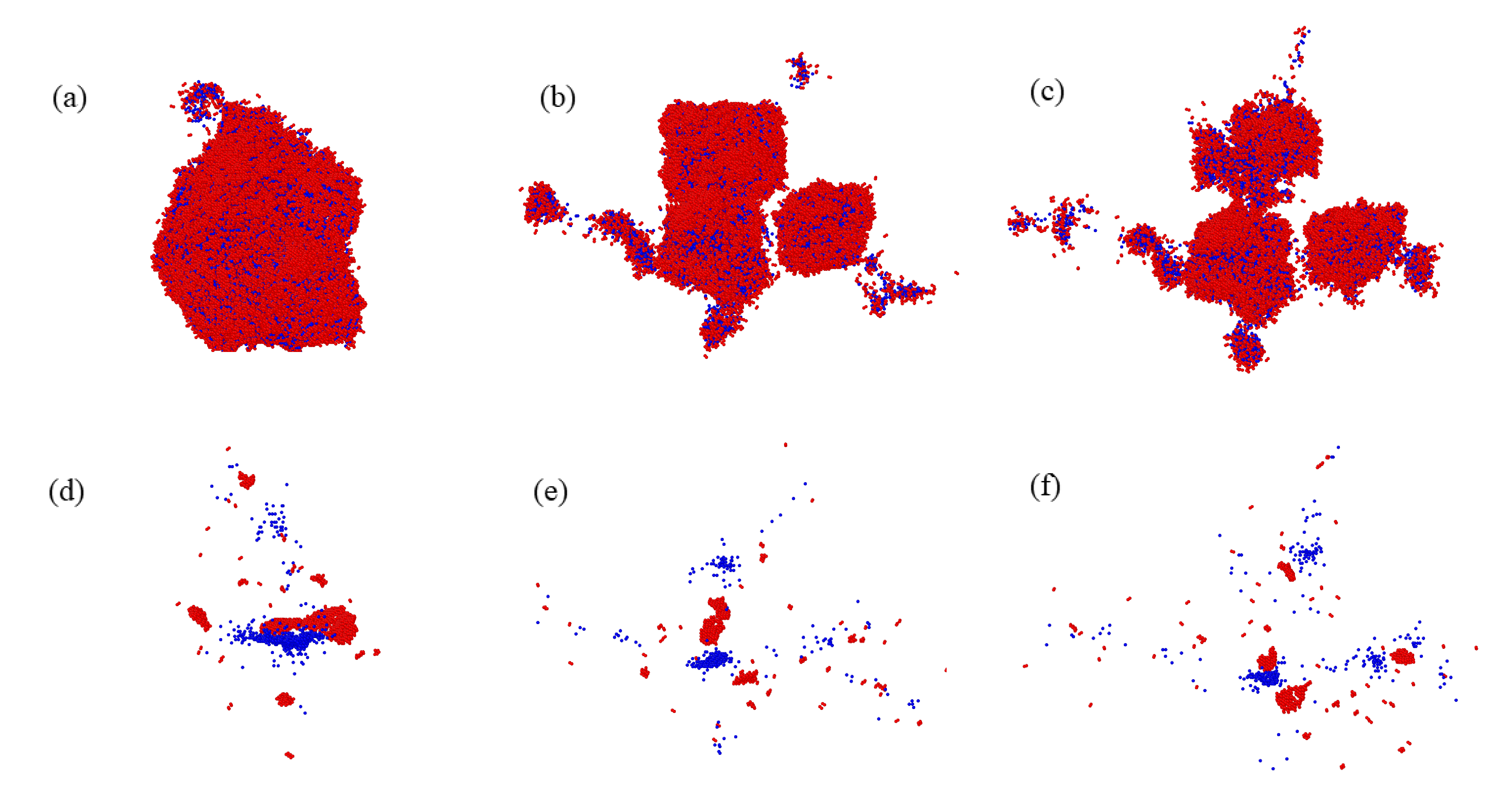}
\caption{Snapshots of three representative cascades at the peak damage states in tungsten: (a) unfragmented type, (b) connected type, and (c) unconnected type. Below them are corresponding snapshots of the surviving defects in the final state of damage: (d) unfragmented type, (e) connected type and (f) unconnected type. The red particles are interstitial atoms and the blue particles are vacancies.}
\label{fig:peak}
\end{figure*}

\begin{figure*}[htb]
\centering
\includegraphics[scale= 0.45]{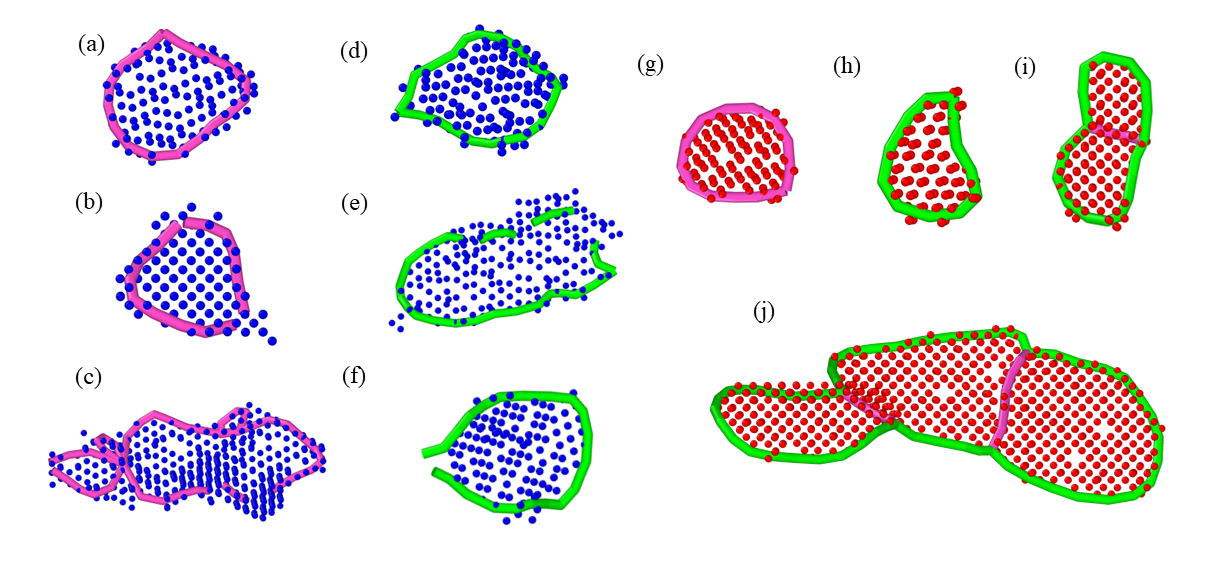}
\caption{(a) A 127-vacancy cluster in form of a complete $\left \langle 100 \right \rangle$ dislocation loop; (b) A 90-vacancy cluster with $\left \langle 100 \right \rangle$ segments; (c) The largest vacancy cluster observed containing 457 vacancies and $\left \langle 100 \right \rangle$ segments; (d) A 104-vacancy cluster in form of a complete 1/2$\left \langle 111 \right \rangle$ dislocation loop; (e) A 203-vacancy cluster with 1/2$\left \langle 111 \right \rangle$ segments, the largest vacancy cluster with this Burgers vector; (f) A 119-vacancy cluster with 1/2$\left \langle 111 \right \rangle$ segments; (g) A complete $\left \langle 100 \right \rangle$ dislocation loop consisting of 55 interstitials; (h) A complete 1/2$\left \langle 111 \right \rangle$ dislocation loop consisting of 40 interstitials; (i) A mixed interstitial loop with a Burgers vectors of 1/2$\left \langle 111 \right \rangle$ and $\left \langle 100 \right \rangle$ consisting of 65 interstitials; (j) The largest interstitial loop observed in W, consisting of 434 interstitials. The red particles are interstitial atoms and the blue particles are vacancies. The green lines represent 1/2$\left \langle 111 \right \rangle$ segments, whereas the pink lines represent $\left \langle 100 \right \rangle$ segments.}
\label{fig:loop}
\end{figure*}

First, we quantified the number of residual point defects in the cascades with \gls{pka} energies ranging from 1 to 200 keV at 300 K. The Norgett-Robinson-Torrens displacements per atom (NRT-dpa) model~\cite{norgett_proposed_1975} is the current international standard for quantifying this energetic particle damage. However, it has been observed that in metals, the number of defects produced in energetic cascades is only one-third of the NRT-dpa prediction \cite{Nordlund2018jnm}. The athermal recombination corrected displacements per atom (arc-dpa) function, proposed by Nordlund et al., improves upon the NRT-dpa by providing a more physically realistic description of primary defect creation in materials \cite{Nordlund2018nc}. Figure~\ref{fig:fks} presents the residual point defects calculated by the \gls{nep}-\gls{zbl} model, the NRT model and the arc-dpa model \cite{Nordlund2018nc}. The value of $E_\mathrm{d}$ used in the model is 90 eV, as commonly used in experimental studies. Setyawan et al. \cite{Setyawan2015jnm} reported two regimes of energy-dependence for defect production in metals. The number of surviving Frenkel pairs (FPs) obtained by the \gls{nep}-\gls{zbl} model is consistent with the arc-dpa model at lower energy regions. In the higher energy region, the number of surviving FPs ($N_F$) follows a function ($N_F=a(E_\mathrm{MD}/E_\mathrm{d})^{b}$) of the reduced energy, $E\equiv E_\mathrm{MD}/E_\mathrm{d}$, with the threshold displacement energy set to 117 eV. The pre-factor is 0.03 and the fitted exponent is 1.25, similar to the results reported by Setyawan et al. \cite{Setyawan2015jnm} with a pre-factor of 0.02 and a fitted exponent of 1.30.

Figure~\ref{fig:peak} presents snapshots of three typical defects resulting from a displacement cascade with a \gls{pka} energy of 200 keV. We classify the cascades by the peak damage state into three categories, unfragmented, unconnected, and connected, following the criterion suggested in Ref.~\onlinecite{Antoshchenkova2015jnm}. The observed probability of subcascade splitting is 70\% for the 200 keV \gls{pka} energy and decreases to 20\% for a \gls{pka} energy of 100 keV. This is consistent with the subcascade splitting threshold for self-ions in tungsten, which is estimated to be around 160 keV based on the analysis of binary collision approximation (BCA) cascades \cite{De_Backer2016epl}. Figure~\ref{fig:peak}~(a) depicts an unfragmented case where the cascade appears in a locally compact region, producing a massive and unbroken molten region, and Figure~\ref{fig:peak}~(d) shows the defect distribution of this case after 150 ps. Large interstitial clusters are accompanied by the formation of large vacancy clusters, both exhibiting a two-dimensional platelet shape (loops). Figure~\ref{fig:peak}~(b)-(e) and Figure~\ref{fig:peak}~(c)-(f) respectively illustrate the continuous morphology of the fragmentations and the defect distribution of them after 150 ps through the connected and the unconnected subcascades. It can be seen that the distribution of defects correlates with the morphology at the peak damage state. Compared to unfragmented cascades, the distribution of defects in fragmented cascades is more dispersed and the clusters are smaller in size.

We analysed each large defect cluster to determine the nature of the dislocation loops. Figure~\ref{fig:loop} shows the typical defect clusters produced by a 200 keV cascade. Most of the dislocation loops observed have the Burgers vector $\bm{b}=1/2\left \langle 111 \right \rangle$. It is worth mentioning that we observed an interstitial $\left \langle 100 \right \rangle$ loop in Figure~\ref{fig:peak}~(f). Although this $\left \langle 100 \right \rangle$ interstitial loop is the only one observed in ten cascades, it is stable within a time scale of 1 ns. The dynamic process of this cascade is showcased in Supplementary Movie 1.

The detailed size distributions of the interstitial and vacancy clusters obtained with 100 keV and 200 keV \gls{pka} energies are shown in Fig.~\ref{fig:cluster_distribution}. The largest clusters of both interstitial and vacancy type were produced by the 200 keV unfragmented cascade. Large vacancy clusters are accompanied by the formation of large interstitial clusters, with sizes more than 80 vacancies mainly exhibiting a two-dimensional platelet shape (loops), while sizes below 80 vacancies exhibit three dimensional shapes (cavity). The interstitial clusters exhibit a two-dimensional platelet shape (loops). Note that a larger number density and smaller average size distribution of defect clusters were found by the 200 keV cascade. With 100 keV \gls{pka} energy, 59\% vacancy clusters are smaller than 50 defects in size, 71\% \gls{sia} clusters are smaller than 100 defects, and the ratios are 79\% and 95\% at 200 keV energy. The more frequent subcascade splitting at 200 keV is the cause of this difference.

\subsection{Radiation-induced defects in thin-foil tungsten}

\begin{figure}[h]
\centering
\includegraphics[width=1\columnwidth]{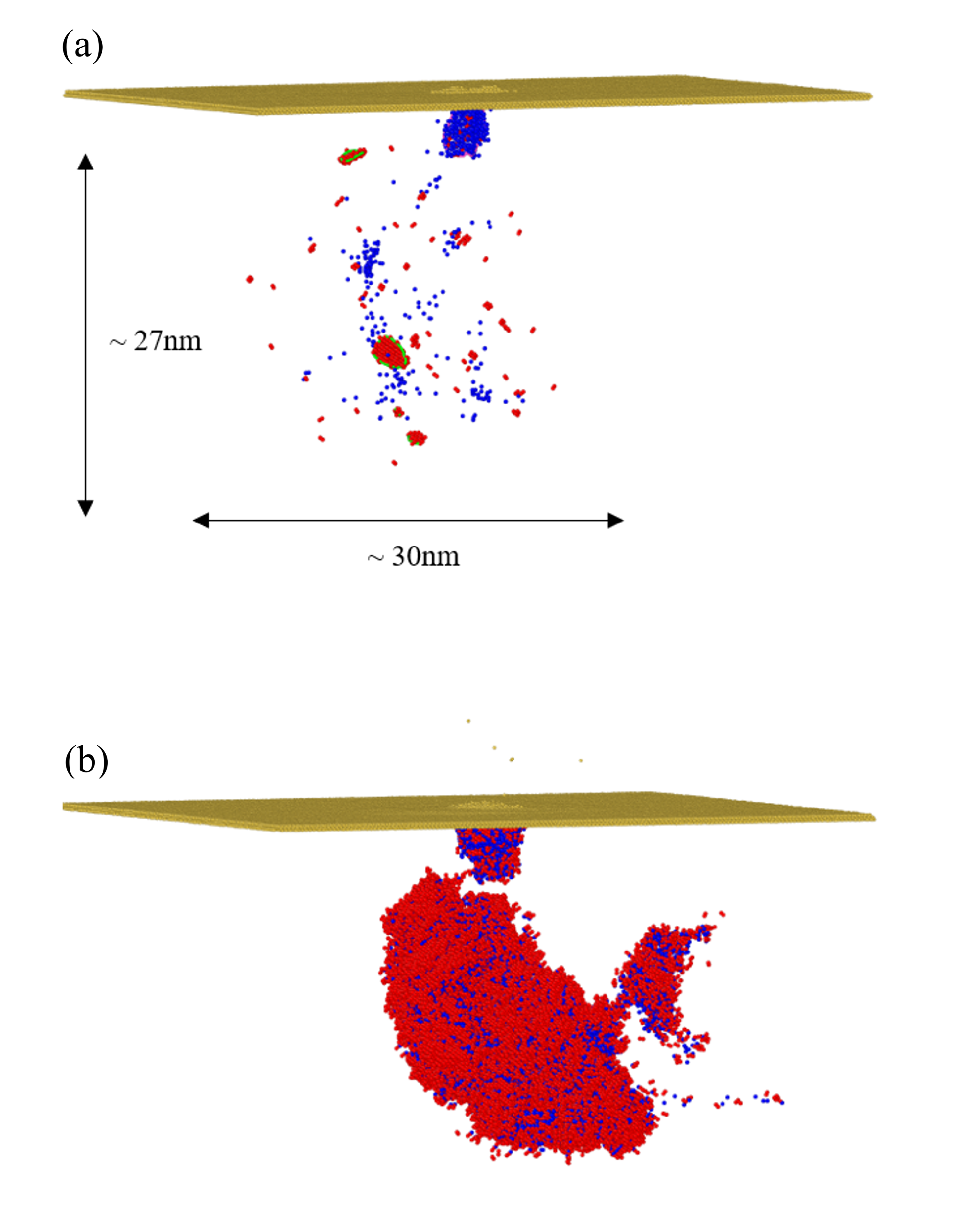}
\caption{Snapshots of surviving defects in the final state of damage~(a) and snapshots of the peak damage state~(b) for cascades on a thin-foil surface. The red particles are interstitial atoms and the blue particles are vacancies. The green lines represent 1/2$\left \langle 111 \right \rangle$ segments, whereas the pink lines represent $\left \langle 100 \right \rangle$ segments.}
\label{fig:surf_damage}
\end{figure}

Figure~\ref{fig:surf_damage} shows the damage microstructure of the tungsten thin film 220 ps after launching the 200 keV \gls{pka} as well as the peak damage state. Blue spheres mark the location of vacancies while red particles are interstitial atoms. Surface-layer atoms, including ad-atoms and sputtered atoms, are also represented in the figure with yellow spheres. In this fragmented cascade, the liquid core of the subcascade heat spike extends to the surface causing near-surface material to be ejected in the form of sputtered atoms and atom clusters. In addition, there is a viscous flow of atoms to the surface resulting in a depleted zone in the core of the cascade and subsequent formation of large vacancy clusters. Figure~\ref{fig:surf_damage}~(a) shows that the damage consists of isolated interstitials and vacancies, three smaller clusters and a relative large cluster of 22, 17, 8, and 76 interstitials, and a significantly larger clusters of 275 vacancies. The total number of vacancies in this case is 478, while the total number of self-interstitials is 212. There are 260 ad-atoms at the free surface and 6 sputtered atoms corresponding to the mismatch between the vacancy and interstitial counts. These observations are consistent with previous simulations of thin-foil tungsten~\cite{sand_defect_2018}.

As can be observed in Fig.~\ref{fig:surf_damage} , most of the damage distribution is scattered, while the displacement cascade reached a maximum depth of 27 nm and width of 30 nm. Figure~\ref{fig:surf_damage}~(a) shows an $\left \langle 100 \right \rangle$ incomplete dislocation loop with 127 vacancies, which remains stable within a time scale of 1 ns. The dynamic process of this cascade is showcased in Supplementary Movie 2. The presence of $\left \langle 100 \right \rangle$ dislocation loops following displacement cascades in W was already shown by Yi et al. \cite{Yi2013pm}. Figure~\ref{fig:surf_damage}~(b) exhibits a snapshot of the cascade at the peak. We find that all cascades with the formation of $\left \langle 100 \right \rangle$ vacancy dislocation loops are fragmented and the subcascade near the surface is unconnected to the main cascade core further below. The $\left \langle 100 \right \rangle$ dislocation loops are formed by cascade collapse from the subscascade. We also obtained one damage configuration which consists of a dislocation network connected to the surface similar to that obtained by Ghaly and Averback in Au \cite{Ghaly1994prl}. This is also consistent with previous simulations of Fe and W thin foils \cite{Aliaga2015acta,sand_defect_2018}.

\begin{figure}[htb]
\centering
\includegraphics[width=\columnwidth]{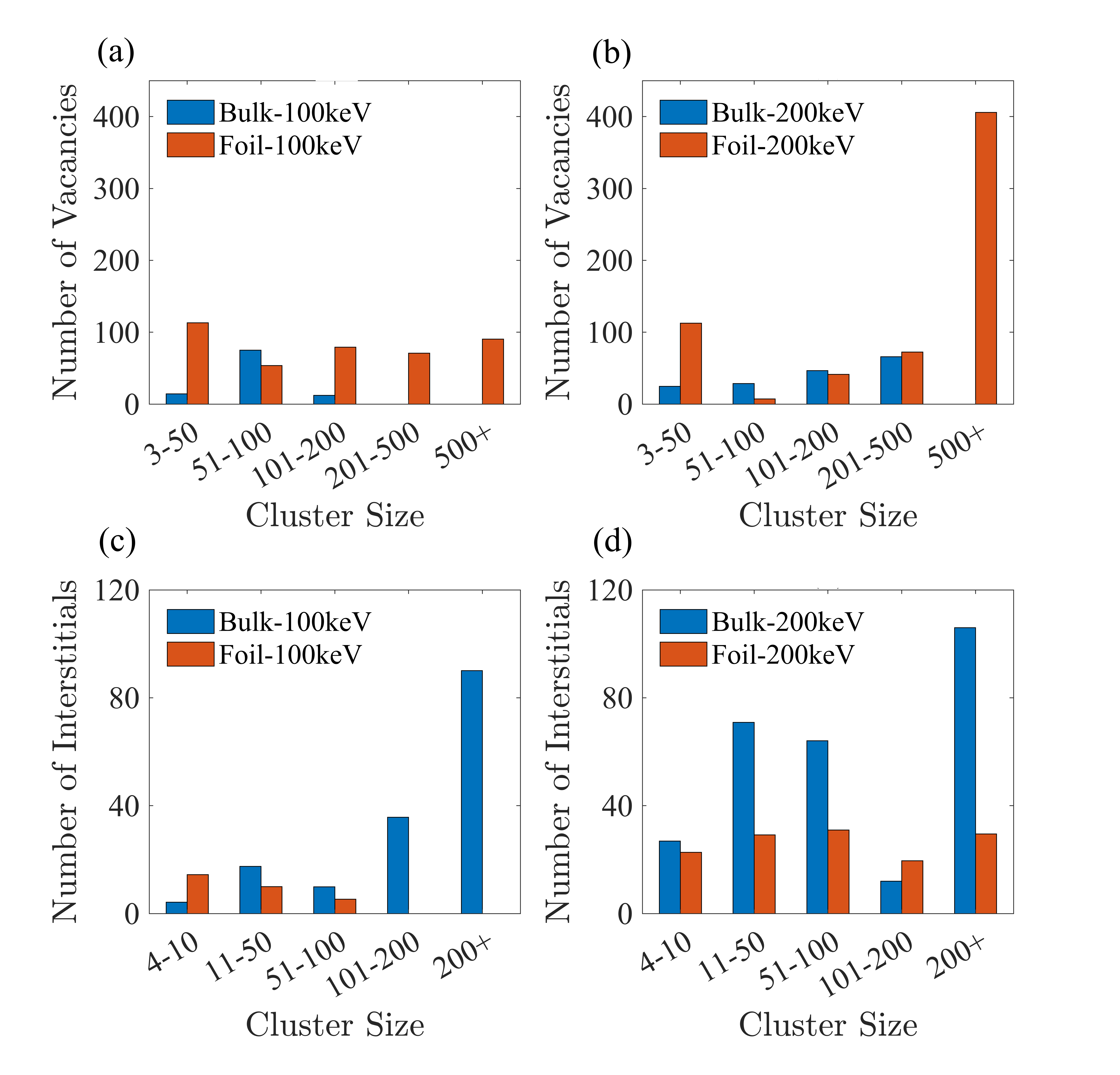}
\caption{Size distribution of defect clusters for cascades simulated in tungsten.}
\label{fig:cluster_distribution}
\end{figure}

In order to quantify the surface effect on radiation-induced defects, we performed a statistical analysis of the data including the mean values of the number of point defects as well as their percentage in clusters, listed in Table~\ref{table:statist}. Due to the surface effect, the number of self-interstitials is always lower than the number of vacancies in foil simulations. As shown in Table~\ref{table:statist} the damage exhibits an increase of 391\% and 264\% for the number of vacancies compared to bulk simulations and a decrease of 66.7\% and 39.2\% for interstitials for 100 keV and 200 keV \glspl{pka}, where the missing self-interstitials correspond to ad-atoms and sputtered atoms \cite{Aliaga2015acta}. The maximum number of atoms sputtered out is 73 with 100 keV \gls{pka} energy. Table~\ref{table:statist} shows that the percentage of vacancies in clusters is very similar for bulk and foil material, but the percentage of interstitials in clusters decreases 37\% and 19\% for 100 keV and 200 keV, respectively. Due to the mechanisms explained above, the main effects of the free surface is that more vacancies are formed but fewer interstitials. Interstitials also remain more isolated instead of efficiently clustering like in bulk cascades. The surface effect decreases with increasing energy since the core of the cascade is further away from the surface with increasing energy.

\begin{table}[h] 
\centering \setlength{\tabcolsep}{1mm} 
\caption{Statistical results of average number of vacancies/interstitials ($N_{\rm{vac/int}}$) and the largest vacancy/interstitial clusters ($S_{\rm{vac/int}}$) following the displacement cascade, and the percentage of vacancies/interstitials in clusters. These data are based on 10 cascade events, with standard deviations in parentheses.}
\begin{tabular}{lllllll}
\hline \hline 
\multicolumn{2}{l}{Energy (keV)}& $N_{\rm{vac}}$&$N_{\rm{int}}$&$S_{\rm{vac}}$&$S_{\rm{int}}$&\% in clusters\\ 
\hline 
100 &Bulk&183(78)&183(78)&97&237&55/86\\ 
&Foil&898(278)&61(72)&905&53&45/49\\ 
200&Bulk&329(142)&329(142)&457&429&50/85\\
&Foil&1200(695)&200(205)&1103&295&52/66\\ 
\hline \hline 
\end{tabular}
\label{table:statist}
\end{table}

Figure~\ref{fig:cluster_distribution} shows the size distributions of clusters of vacancies and interstitials in thin foils. It is clear that larger and more vacancy clusters as well as smaller and less interstitial clusters are produced, compared to bulk tungsten. In addition, the scatter in the size of clusters is extensive. There is one vacancy cluster with 1103 vacancies formed by a 200 keV cascade. The surface also affects the dislocation density. Compared to bulk, the dislocation density of $\left \langle 100 \right \rangle$ Burgers vectors increases from $9.88\times10^9~\rm{cm}^{-2}$ to  $1.08\times10^{11}~\rm{cm}^{-2}$ as well as from $4.47\times10^{10}~\rm{cm}^{-2}$ to $9.86\times10^{10}~\rm{cm}^{-2}$ for 100 keV and 200 keV \glspl{pka} respectively. Correspondingly the 1/2$\left \langle 111 \right \rangle$ dislocation density decreases from $2.87\times10^{11}~\rm{cm}^{-2}$ to  $2.01\times10^{11}~\rm{cm}^{-2}$ as well as from $2.82\times10^{11}~\rm{cm}^{-2}$ to  $1.69\times10^{11}~\rm{cm}^{-2}$. Overall, the simulations show that the frequency and size of vacancy dislocation loops and the $\left \langle 100 \right \rangle$-dislocation density is greater when damage is produced in thin foils.

In our computational modeling study, we have observed results consistent with the experimental findings reported by Yi \textit{et al.}~\cite{Yi2015acta}. The emergence of 1/2$\left \langle 111 \right \rangle$ and $\left \langle 100 \right \rangle$ vacancy loops in low-dose heavy-ion irradiated tungsten is validated as an intrinsic cascade phenomenon. This suggests that these loops primarily originate from cascade nucleation rather than resulting from long-term evolution of the defect structure. Moreover, our simulations emphasize the impact of foil surfaces, as seen in thin-foil irradiation specimens or back-thinned irradiated specimens, on the evolution of damage. The notable trends elucidated by Yi \textit{et al.}~\cite{Yi2016acta} were consistently replicated in our computational investigation: the total vacancy count was observed to be an order of magnitude greater than the number of \glspl{sia}, indicating a substantial effect of proximate free surfaces on defect formation. In contrast to EAM models wherein the effect of surface on defect statistics is more pronounced in Fe than in W \cite{Sand2016epl} and the formation of SIA clusters is only slightly affected by the surface, the NEP-ZBL potential shows a noticeable impact of the free surface in tungsten. This difference suggests that EAM models may not fully capture free surface effects, which is in line with the fact that surfaces in EAM models are often significantly too stable (i.e., surface energies are severely underestimated, Fig.~\ref{fig:surf}). 

\section{Summary and Conclusions}

We have introduced a hybrid scheme of the Ziegler-Biersack-Littmark (\gls{zbl}) screened nuclear repulsion potential and the neuroevolution machine learning potential. This novel model achieves accuracy comparable to other ML potentials, while exhibiting high computational efficiency in terms of computation time and memory usage. This allowed us to investigate energetic radiation-induced collision cascades in large-scale molecular dynamics simulations. Primary radiation damage of tungsten was investigated by irradiation with ions of energies ranging from 1 keV to 200 keV in both thin foil and bulk form. We observed that low-dose high-energy irradiation directly generates $\left \langle100\right \rangle$ interstitial dislocation loops in tungsten, which remained stable within 1 ns. The simulations also generated complete dislocation loops of vacancy type, including $\left \langle100\right \rangle$ dislocation loops and 1/2 $\left \langle111\right \rangle$ dislocation loops. The presence of a surface led to the formation of more numerous and larger vacancy clusters as well as smaller and less interstitial clusters. Some of vacancy clusters coalesced into incomplete $\left \langle100\right \rangle$ dislocation loops. This significantly enhances the linear density of $\left \langle100\right \rangle$ dislocations in the irradiated material. Our results shows that one should carefully account for the effect of free surfaces in tungsten. 

We showed that the \gls{nep}-\gls{zbl} potential provides accurate predictions of material properties which have been persistent challenges for empirical potentials, such as the relative stability of defect clusters and surface properties. Furthermore molecular dynamics simulations revealed that our findings for bulk tungsten agree with existing results from EAM models. However we see pronounced differences in the simulations for foils compared to EAM models. These differences suggest that while EAM models can depict radiation damage in bulk structures, they may not fully reproduce the effects of free surfaces. With an efficient \textsc{gpu} implementation, the \gls{nep}-\gls{zbl} framework hence offers good opportunities for large-scale simulations of radiation damage also in other materials, particularly in systems lacking suitable empirical potentials such as high-entropy alloys and ceramics.

\textbf{Data availability}

The training and validation results for the NEP-ZBL model of tungsten can be freely available from the Github repository \url{https://github.com/Jonsnow-willow/GPUMD-Wizard}. Other data presented in this paper are available from the corresponding authors upon reasonable request.

\begin{acknowledgments}
The authors acknowledge funding from the CNNC Science Fund for Talented Young Scholars FY222506000902. (Y. S. \& J. L.), the National Natural Science Foundation of China (NSFC) under grant no. 11974059 (Z. F.), the Academy of Finland Flagship programme: Finnish Center for Artificial Intelligence FCAI (J. B.), the National Key Research and Development Program of China under grant no. 2021YFB3802104 (P. Q.), and USTB MatCom of Beijing Advanced Innovation Center for Materials Genome Engineering. 
\end{acknowledgments}

\bibliography{ref.bib}

\end{document}